\documentclass[12pt]{article}
\usepackage{amssymb,amsmath,bm,mathrsfs,subfigure,graphicx}
\begin{document}
\begin{center}
\begin{Large}
{\bf  How quantum mechanics probes superspace}
\end{Large}
\vskip1truecm
Stam Nicolis\footnote{E-Mail: Stam.Nicolis@lmpt.univ-tours.fr}

{\sl CNRS--Laboratoire de Math\'ematiques et Physique Th\'eorique (UMR 7350)\\
F\'ed\'eration de Recherche ``Denis Poisson'' (FR 2964)\\
D\'epartement de Physique\\
Universit\'e ``Fran\c{c}ois Rabelais'' de Tours\\
Parc Grandmont, Tours 37200, France}

\end{center}
\begin{abstract}
 We study the relation between the partition function of a non--relativistic particle,  that describes the equilibrium fluctuations implicitly, and the partition function of the same system, deduced from the Langevin equation, that describes the fluctuations explicitly, of a bath with additive white--noise properties.  We show that both can be related to the partition function of an ${\mathcal N}=1$ supersymmetric theory with one--dimensional bosonic worldvolume and that they can all describe the same physics, since the correlation functions of the observables satisfy the same identities for all systems.The supersymmetric theory provides the consistent closure for describing the fluctuations, even though supersymmetry may be broken, when their backreaction is taken into account. The trajectory of the classical particle becomes a component of a superfield, when fluctuations are taken into account. These statements can be tested by the identities the correlation functions satisfy, by using a lattice regularization of an action that describes commuting fields only.

PACS: 02.50Ey, 02.70Uu, 03.65Ca, 05.10Gg, 11.15Tk, 11.30Pb.
\end{abstract}
The stochastic approach for describing the  dynamics of commuting degrees of freedom starts with the Langevin equation in the presence of, additive, white noise
\begin{equation}
\label{langevin1}
\frac{\partial\phi(t)}{\partial t} = -\frac{\partial U(\phi)}{\partial\phi(t)}
+ \eta(t)
\end{equation}
where $t$ is the equilibration time. The field, $\eta(t)$, is a Gaussian
stochastic process:
\begin{equation}
\label{gaussian_noise}
\begin{array}{l}
\left\langle\eta(tà\right\rangle = 0\\
\left\langle\eta(t)\eta(t')\right\rangle = \nu\delta(t-t')\\
\left\langle
\eta(t_1)\eta(t_2)\cdots\eta(t_{2n})
\right\rangle
= \sum_{\pi}
\left\langle \eta(t_{\pi(1)})\eta(t_{\pi(2)})\right\rangle\cdots
\left\langle \eta(t_{\pi(2n-1)})\eta(t_{\pi(2n)})\right\rangle
\end{array}
\end{equation}
where the sum is over all permutations. The qualification ``additive'' means that the coefficient of the noise term does not depend on the dynamical variable(s), $\phi$ and can be taken as a constant, whose value can be set to 1.

Depending on the interpretation given to this coefficient, the fluctuations are quantum, thermal, or due to disorder, if the coefficient is identified with $\hbar$, $k_\mathrm{B}T$ and the strength of the disorder respectively. We shall use language appropriate for the quantum case, but the formalism is general.

If $U(\phi)$ is a local functional of $\phi$, in particular, if 
\begin{equation}
\label{Uqm}
\frac{\partial U(\phi)}{\partial\phi} =\frac{\partial\phi(\tau)}{\partial\tau}
+ \frac{\partial W(\phi)}{\partial\phi(\tau)}
\end{equation}
where $\tau\in\mathbb{R}$ and $W(\phi(\tau))$ an ultralocal functional of
$\phi(\tau)$, we obtain the following stochastic equation for $\phi(\tau)$:
\begin{equation}
\label{langevin2}
\frac{\partial\phi(\tau)}{\partial\tau} + \frac{\partial W}{\partial\phi(\tau)}=\eta(\tau)
\end{equation}
This is the Langevin equation that describes, for example, quantum mechanics, i.e. a quantum
field theory in one Euclidian dimension. The essential difference to
eq.~(\ref{langevin1}) is that we are not interested, only, in the limit
$\tau\to\infty$, but in  the solution for all values of $\tau$. This assertion
is meaningful only at the level of the correlation functions, of course.

In this case, $\eta(\tau)$ is a Gaussian stochastic process, whose correlation
functions obey the same identities as in eq.~(\ref{gaussian_noise}), only the
time is, now, the Euclidian time. 

We are interested in the correlation functions,
$\left\langle\phi(\tau_1)\phi(\tau_2)\cdots\phi(\tau_n)\right\rangle$, of the
field $\phi$ and the identities that constrain them. This information can be obtained from the partition function of the noise, that describes the properties of the correlation functions of the noise and the Langevin equation, that defines the change of variables from the noise, $\left\{\eta(\tau)\right\}$ to the field $\{\phi(\tau)\}$~\cite{parisi_sourlas,nicolai}.

What we wish to point out is that there are three partition functions that can be used to define these correlation functions:
\begin{equation}
\label{threeZ}
\begin{array}{l}
\displaystyle
Z_\mathrm{L}=\int\,\left[{\mathscr D}\phi\right]\,\mathrm{det}\left(\frac{d}{d\tau}+W''(\phi)\right)\mathrm{sign}\left(\mathrm{det}\left(\frac{d}{d\tau}+W''(\phi)\right) \right)
\,e^{-\int\,d\tau\,\frac{1}{2}\dot{\phi}^2+V(\phi)}\\
\displaystyle
Z_\mathrm{QM}=\int\,\left[{\mathscr D}\phi\right]\,e^{-\int\,d\tau\,\frac{1}{2}\left(\dot{\phi}+W'(\phi)\right)^2}
\\
\displaystyle
Z_\mathrm{SUSY}=\int\,[{\mathscr D}\phi][{\mathscr D}\psi][{\mathscr D}\chi][{\mathscr D}F]\, e^{-S[\phi,\psi,\chi,F]}
\\
\displaystyle 
S[\phi,\psi,\chi,F]=\int\,d\tau\,\left\{
 \frac{1}{2}\dot{\phi}^2-\frac{F^2}{2}+FW'(\phi)-\frac{1}{2}\left(\psi\dot{\chi}-\dot{\psi}\chi\right)-\frac{1}{2}W''(\phi)\left[\psi,\chi\right]
 \right\} 
\end{array}
\end{equation}
where $V(\phi)=(1/2)(W'(\phi))^2$ up to a constant and periodic boundary conditions imply that the mixed term, $\dot{\phi}W'(\phi)$, which is a total derivative, doesn't contribute. The periodic boundary conditions can be shown to imply that $Z_\mathrm{SUSY}$ does, in fact, compute the Witten index~\cite{alvarez-gaume}.

The auxiliary field, $F(\tau)$ is introduced~\cite{dAFFV} in order to realize the transformations that map commuting to anticommuting fields and have anticommuting parameters linearly in the fields. Since there is considerable variation in the conventions, that can lead to confusion, we provide an example of the transformations, 
\begin{equation}
\label{SUSYalgebra}
\begin{array}{lcl}
\displaystyle
Q_1\phi = -\chi & & \displaystyle Q_2\phi = \psi\\
\displaystyle
Q_1\chi = 0 & &\displaystyle Q_2\chi = \dot{\phi}-F\\
\displaystyle 
Q_1\psi = -\dot{\phi}-F & &\displaystyle Q_2\psi = 0\\
\displaystyle 
Q_1 F = \dot{\chi} & & \displaystyle Q_2 F = \dot{\psi}
\end{array}
\end{equation}
that leave $S_\mathrm{SUSY}$  invariant, $\delta_1 S=\zeta_1Q_1 S=0=\delta_2 S =\zeta_2 Q_2 S$, up to total derivatives,  that are nilpotent, $Q_1^2=0=Q_2^2$ and whose anticommutator closes on the generator of translations, in Euclidian time, $\{Q_1,Q_2\}=-2(d/d\tau)\Leftrightarrow \left[\zeta_1Q_1,\zeta_2Q_2\right]=2\zeta_1\zeta_2d/d\tau=-\zeta_\alpha\varepsilon^{\alpha\beta}\zeta_\beta d/d\tau$.

They all describe the same physical system, because they represent three different ways of calculating the same path integral, that of the noise,
\begin{equation}
\label{noise}
Z=\int\,[{\mathscr D}\eta(\tau)]\,e^{-\int\,d\tau\frac{1}{2}\eta(\tau)^2}=1
\end{equation}
The normalization can be chosen so that this partition function is equal to a universal constant, independent of the parameters of the distribution and, thus, may be set to unity. The statement that, in fact, all three can be set equal to a universal constant, that doesn't depend on the dynamics, is an expression of the fact that the system is consistently closed.

\begin{figure}[thp]
\label{2point}
\begin{center}
\subfigure{\includegraphics[scale=0.6]{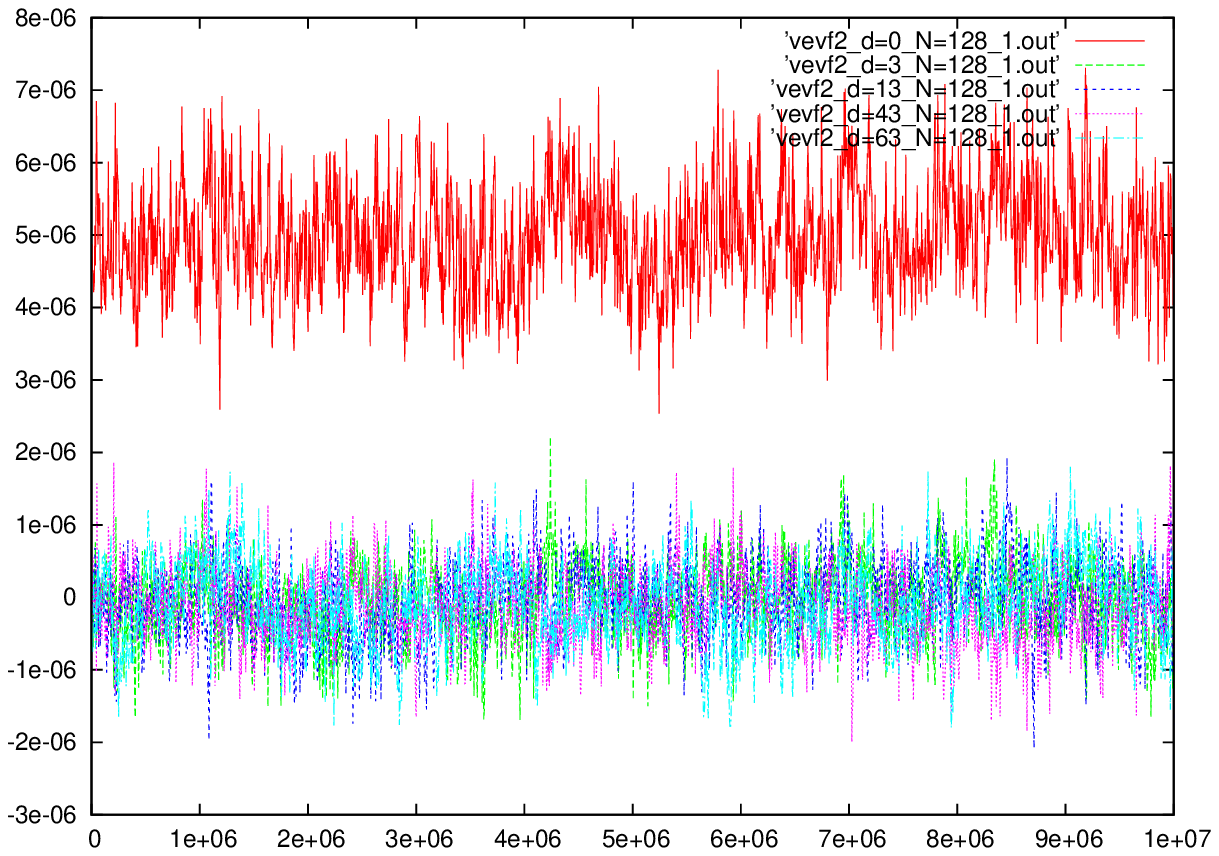}}
\subfigure{\includegraphics[scale=0.6]{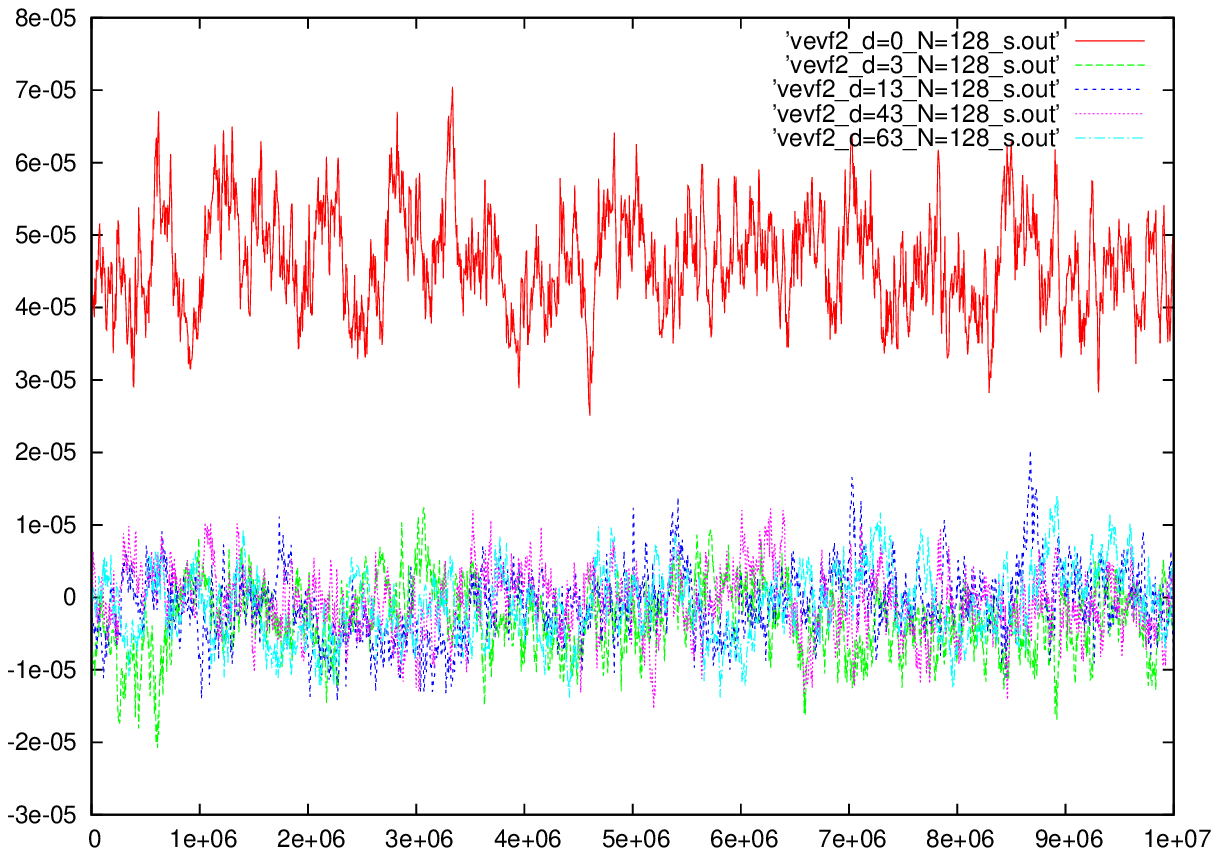}}
\caption[]{Monte Carlo time series, for the 2--point,  functions, $\langle\mathfrak{h}_{|n-n'|}\mathfrak{h}_0\rangle$,  of the noise field, for $|n-n'|<64$, for $N=128$, $g=0.1$ and  $g=1.0$.   The value of $m_\mathrm{latt}^2=0.0001$.}
\end{center}
\end{figure}

To test that these three partition functions are equivalent, we use a lattice regularization of $Z_\mathrm{QM}$, 
\begin{equation}
\label{ZQMlatt}
\begin{array}{l}
\displaystyle Z_\mathrm{QM}^\mathrm{latt}=\int\,\left[\prod_{n=0}^{N-1}\,d\Phi_n\right]\,e^{-S_\mathrm{latt}[\Phi_n]}\\
\displaystyle
S_\mathrm{latt}=\frac{1}{gm_\mathrm{latt}^2}\sum_{n=0}^{N-1}\left[ -\Phi_n\Phi_{n+1}+\Phi_n^2 + \frac{m_\mathrm{latt}^4}{2}\left(\Phi_n+\frac{\Phi_n^3}{6}\right)^2\right]
\end{array}
\end{equation}
to define the measure  and the action and compute the appropriately regularized correlation functions of the noise field, 
\begin{equation}
\label{lattnoise}
\mathfrak{h}_n=\frac{1}{2}\left(\Phi_{n+1}-\Phi_{n-1}\right) +m_\mathrm{latt}^2\left(\Phi_n+\frac{\Phi_n^3}{6}\right)
\end{equation}
(for the case of the quartic superpotential with a unique minimum)
 by Monte Carlo sampling of an action that is well--defined, since it's bounded from below, and contains only the commuting field, $\phi(\tau)$. It is the fact that the particular combination of the scalar field defined by eq.~(\ref{langevin2}), and sampled by the action of $Z_\mathrm{QM}$, can be shown to define a Gaussian field, with ultra--local propagator, that expresses this equivalence. The lattice regularization tests the irrelevance of surface terms and terms that are proportional to positive powers of the lattice spacing.  It also tests that the sign of the determinant is correctly taken into account, albeit implicitly. A sample of the numerical results is presented in fig.~\ref{2point} that shows that the 2--point function of the noise field on the lattice is, in fact, a $\delta-$function, for weak ($g=0.1$) as well as for strong ($g=1.0$) coupling.

A detailed presentation of results of the simulations and a fuller discussion is in preparation in an updated version of ref. ~\cite{nicolis}.

In conclusion we have shown how the stochastic approach for describing the equilibrium fluctuations of a non--relativistic particle with one--dimensional target space can quite naturally be encoded by the properties of a ${\mathcal N}=1$ supersymmetric theory. The supersymmetry is worldline supersymmetry and implies that the worldvolume, here the worldline, $\tau$, becomes a supermanifold, $(\tau,\theta,\overline{\theta})$ and the target space, parametrized by the scalar field, $\phi$, also, becomes a supermanifold, $(\phi,\psi,\chi,F)$. What the backreaction implies is that it isn't possible to integrate out the anticommuting fields at fixed backrground for the commuting fields~\cite{witten}.

It is, of course, natural to study how boundary conditions (the boundary degrees of freedom were studied in ref.~\cite{nicolis_zerkak}) and constraints~\cite{constraints}, especially of the target space, can be taken into account, when the backreaction of the fluctuations cannot be neglected and, further, how target space supersymmetry can be described. This is work in progress and we look forward to reporting on it in forthcoming publications.

{\bf Acknowledgements}: It's a pleasure to thank the organizers, S. Fedoruk,  E. Ivanov and S. Sidorov  of the SQS15 Workshop for the warm hospitality, and the opportunity for inspiring discussions. I would like to particularly thank C. P. Bachas, E. Ivanov and A. Smilga for their interest in this work and their  constructive criticism.

\end{document}